# Theory and MD simulations of intrinsic localized modes and defect formation in solids

V. Hizhnyakov[1], M. Haas[1], A. Shelkan[1] and M. Klopov[2]

[1] Institute of Physics, University of Tartu, Riia 142, 51014 Tartu, Estonia
[2] Tallinn University of Technology, Ehitajate tee 5, 19086 Tallinn, Estonia

E-mail: hizh@fi.tartu.ee

**Abstract.** MD simulations of recoil processes following the scattering of X-rays or neutrons have been performed in ionic crystals and metals. At small energies (<10 eV) the recoil can induce intrinsic localized modes (ILMs) and linear local modes associated with them. As a rule, the frequencies of such modes are located in the gaps of the phonon spectrum. However, in metallic Ni, Nb and Fe, due to the renormalization of atomic interactions by free electrons, the frequencies mentioned are found to be positioned above the phonon spectrum. It has been shown that these ILMs are highly mobile and can efficiently transfer a concentrated vibrational energy to large distances along crystallographic directions. If the recoil energy exceeds tens of eVs, vacancies and interstitials can be formed, being strongly dependent on the direction of the recoil momentum. In NaCl-type lattices the recoil in (110) direction can produce a vacancy and a crowdion, while in the case of a recoil in (100) and in (111) directions a bi-vacancy and a crowdion can be formed.

## 1. Introduction

A number of important for various applications phenomena in condensed matter are govern by atomic (ionic) motions which are characterized by large displacements and strong nonlinear forces. As an important example may serve the radiation damage due to production of defects at the irradiation of solids by neutrons, X-rays and other high energy particles in fusion and nuclear reactors. Here in addition to initial, high energy cascade damage, the final result is essentially determined also by the final stage of the damage which results from the collisions with particles of low-energy. In this communication we investigate this final stage characterized by the moderate (0.1 eV < $E_R$ < 200 eV) recoil energy of collision $E_R$. In this energy range, if $E_R$ < 10 eV (i.e.at the atomic velocities smaller than the phonon phase velocities), one can expect the creation of intrinsic localized modes (ILMs, also known as vibrational solitons, breathers, quodons) [1-3]. An analytical theory of ILM (see also Refs.[4,5]) was developed and the possible trapping of phonons by ILM was predicted. The corresponding linear local modes (LLMs) manifest themselves as a modulation of ILMs amplitudes [6]. Recently, the existence of LLMs has been confirmed experimentally [7]. To investigate the localization and the transport of vibrational energy in a form of immobile and mobile ILM, numerical simulations were performed by us in metals. As a specific character, in metals ILMs may exist only if their frequency is positioned above the phonon spectrum, which assumes atomic interactions with rather small odd anharmonicity. Our simulations showed that in some metals (Ni, Nb, Fe) such high-frequency ILMs are possible due to the screening of the short-range ion-ion interaction by the conducting electrons accompanied by a strong suppression of odd anharmonicity [8]. We have found that these excitations may have rather large energy and they can move without any remarkable decay over large distances along crystallographic directions in the lattice.

We also studied the evolution of the lattice excitations induced by the recoil processes following the scattering of X-rays or neutrons in crystals at recoil energies 10 eV < $E_R$ < 200 eV, i. e. in the range where the vacancies and interstitials can be formed. We numerically simulated the formation of these defects in ionic crystals in dependence of energy, direction of the recoil momentum and of the atomic (ionic) masses of the crystal. In particular, it was shown that the defect pair, bivacancy+crowdion, can arise in NaI as a result of the great anion-cation mass difference. The long-range forces were partially taken into account in the simulations and we found that they can affect the behaviour of the crowdion-type defect created at recoil.

## 2. Self- consistent theory of ILMs and LLMs

In Ref. [4] a theory of ILMs is proposed, which is based on the consideration of small deviations. The latter satisfy the self-consistent linear equation $\ddot{\alpha}_n = -\sum_{n'}(D_{nn'} + w_{nn'})\alpha_{n'}$. Here $D_{nn'}$ is the dynamical matrix of the harmonic lattice and $w_{nn'}$ is its perturbation by an ILM. If to take an ILM in the form $u_L(t) = A_n \cos(\omega_L t) + O$ (the term $O$ describes small higher-order harmonics), then the self-consistency condition reads $\alpha_n = \zeta \cdot A_n \sin \omega_L t$, where $\zeta$ is a small parameter (note the phase shift $\pi/2$ of the deviations with respect to the ILM). For such $\alpha_n$ one gets [4]

$$w_{nn'} = 2\langle \sin^2 \omega_L t\, \partial^2 V_{anh}/\partial u_n \partial u_{n'} \rangle, \qquad (1)$$

where $\langle \cdots \rangle$ is the averaging over the vibrational period (the explicit equations for $w_{nn'}$ are given in [4]). The solution of the equation for $\alpha_n$ reads $\alpha_n = \alpha_0 \tilde{G}_{n0}/\tilde{G}_{00}$, where

$$\tilde{G}(\omega) = (I - G(\omega)w)^{-1} G(\omega) \qquad (2)$$

$G(\omega)$ is the Green's function of phonons in its spectral representation.

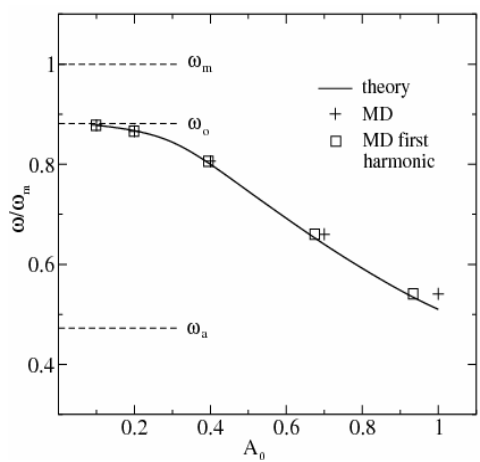
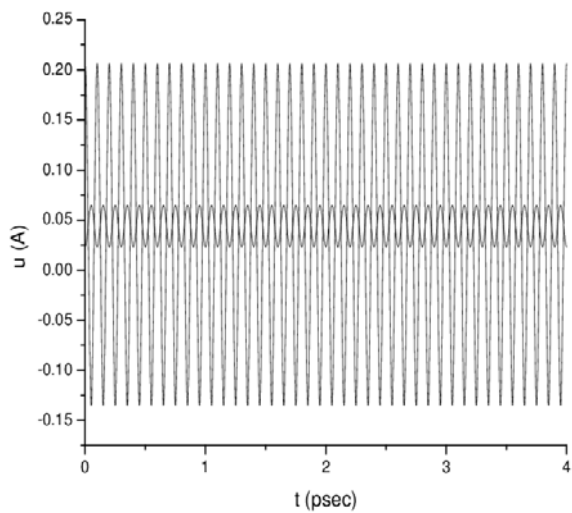

**Figure 1**. Frequency of ILM in NaI chain versus amplitude of the central Na ion; $\omega_m$ and $\omega_o$ are the top and the bottom of the optical band, $\omega_a$ is the top and the bottom of the acoustic band.

**Figure 2.** An immobile even ILM in Fe: the time dependence of the oscillations of two atomic bonds: central and the third from the centre (with small amplitude).

Equation (2) is a self-consistent equation which can be solved selecting $A_n$ and $\omega_L$, step by step. The theory has been checked by numerical simulations of ILMs in monatomic and two-atomic chains [4,5] (see figure 1) and it has been applied for alkali-halide crystals [9]. Analogously, one can also consider the effect of an ILM on the rest of the small vibrations (phonons) in the lattice [6].

## 3. Mobile ILMs with the frequency above the phonon spectrum

In Ref. [8] we derived a critical parameter for the ILMs existence: $\bar{k} = 3k_4\bar{k}_2/4k_3^2$ (here $k_3$ and $k_4$ are the cubic and quartic anharmonic springs, $\bar{k}_2 = Mv^2/r_0^2$ is the mean elastic spring in the bulk crystal, $v$ is the longitudinal velocity of sound, $M$ is the mass of atoms, $r_0$ is the lattice constant). In insulators all standard pair-potentials, such as Morse, Lennard-Jones, Toda, Buckingham, Born-Mayer+Coulomb, used at the MD simulations of ILMs, show a strong softening at the increasing of vibrational amplitudes ($\bar{k} < 1$) due to large odd anharmonicity. As a result, the possible ILMs can only split down from the optical band(s) of the phonon spectrum into the spectral gaps. In metals, in the absence of the gaps mentioned the situation is very different. Here the condition $\bar{k} > 1$ must be fulfilled to enable the position of a possible ILM above the phonon spectrum. We have shown that at least in some metals (Ni, Nb, Fe) $\bar{k} > 1$, as the odd anharmonicity is strongly reduced due to the screening of the ion-ion interaction by the conducting electrons, and ILMs can exist. This conclusion was verified by the MD simulations of the recoil-induced dynamics in these crystals.

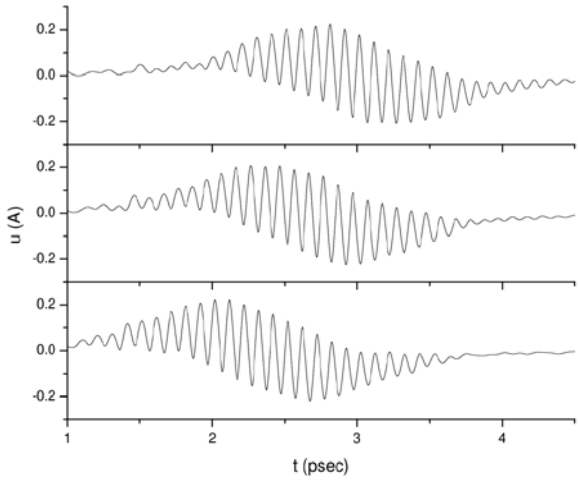 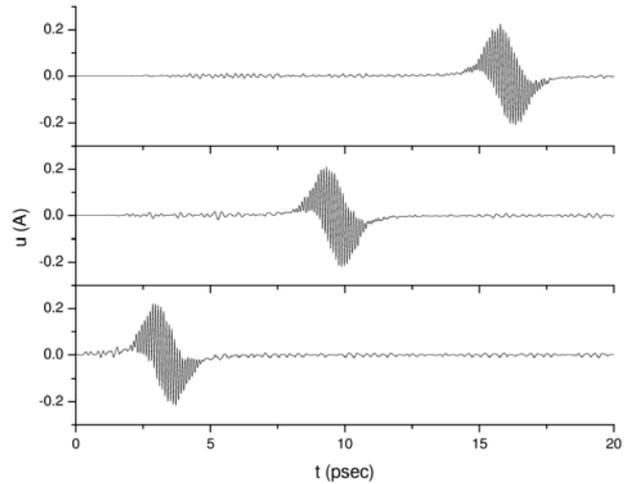

**Figure 3.** Mobile ILM in Fe: the time dependence of the vibrations of the atoms No. 7 (lower curve), No. 8 (middle curve) and No. 9 (upper curve) in the (111) chain (the initial excitation at the atom No. 1).

**Figure 4.** Mobile ILM in Fe: the time dependence of the vibrations of the atoms No. 10 (lower curve), No. 30 (middle curve) and No. 50 (upper curve) in the (111) chain (the initial excitation at the atom No. 1).

Very stable and well mobile ILMs have been found in iron (the mobility of the ILM is determined by the momentum of the initial excitation). In our calculations we have used the latest EAM potential found by Mishin et al [11,12]. Two series of calculations were performed: 1) using the original program and 2) applying a LAMMPS package. Both calculations fully agree with each other. In the case of the immobile ILM (figure 2) and small motion of ILM (figure 3) the cluster with $40 \times 40 \times 40$ iron atom cells was used for MD simulations; for an extended motion (figure 4) cluster $12 \times 12 \times 12$, prolonged in the (111) direction to have 200 atoms in the (111) chain, was used (total number of atoms 73328). The full energy of the ILM, represented in figures 3 and 4, is 1.5 eV.

## 4. MD simulations of recoil-induced defects

We have also studied the recoil-induced dynamics of NaI and KCl crystals at recoil energies $E_R > 10$ eV. In this energy range the generation of vacancies and interstitials can be expected instead of ILMs. We performed MD simulations in the rectangular clusters, extended along the crystallographic axes (the main directions of the studied recoil momenta). Clusters contain 3000 atoms and have the length of $60a_0$ in the (100) direction or $60a_1$ ($a_1 = \sqrt{2}\,a_0$) in the (110) direction ($a_0$ is the lattice parameter). The clusters were located in the infinite host lattice of immobile ions. In the cluster, direct integration of the equations of the atomic motion was carried out by using the short-range forces (Born-Mayer + van der Waals) of Ref. [13], cut off at $8$ Å $< r < 9$ Å. Coulomb forces between all ions of the cluster were taken into account and the Coulomb interaction with the host (anharmonic forces) was calculated by using the Ewald method. The approximation of nonpolarizable ions was used in simulations. Note that the present potentials differ from the model used in Ref. [10]. The latter generated unrealistic results at some recoil momentum directions. The change of the model potential decreases also the threshold of the defect formation in the (110) direction case.

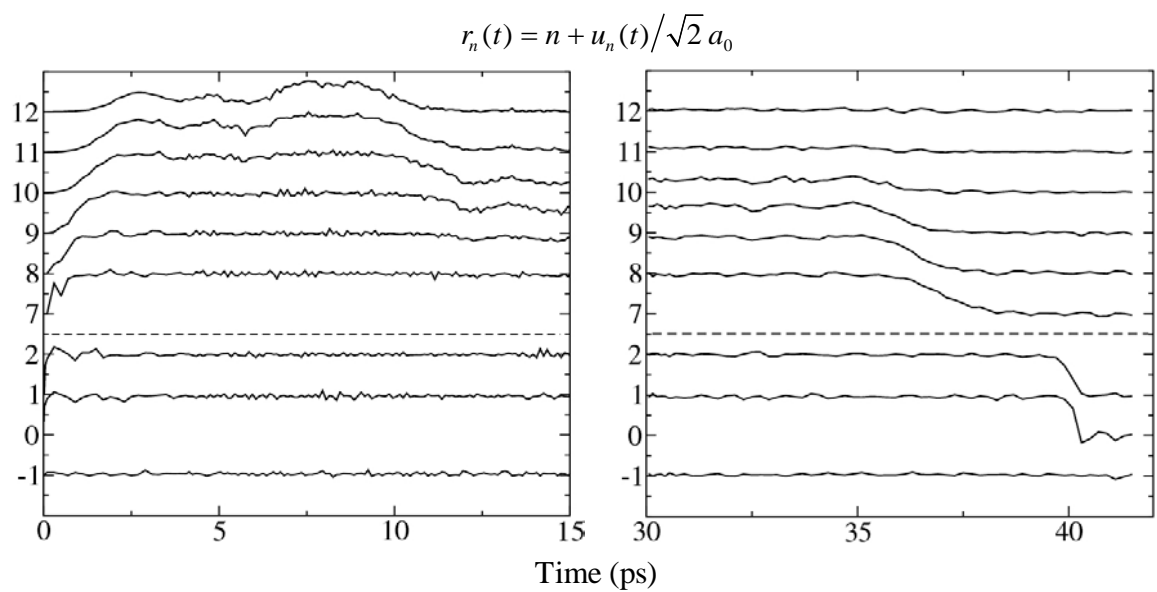

$$r_n(t) = n + u_n(t)/\sqrt{2}\,a_0$$

**Figure 5.** Time dependence of the positions of the colliding ions in a NaI crystal. Displacements are induced by a recoil momentum transferred to the primary Na ion ($n = 0$) in the (110) direction. Recoil energy $E_R = 27$ eV, lattice parameter $a_0 = 3.208$ Å.

The lowest defect creation threshold (recoil energy $E_R \approx 25$ eV) was found for the (110) direction of the recoil. In such direction the recoil causes a cascade of collisions in the (110) chain of identical atoms. As a result, one can observe the creation of a vacancy and a crowdion in the chain (figure 5), the crowdion-vacancy distance increasing with $E_R$. At exact (110) direction, an intensive crowdion drift towards vacancy took place after a 35 ps interval, which caused the annihilation of the defect pair. At deviations of the recoil direction from the (110) direction, the momentum transfer was accompanied by strong focuson effects in the solid angles $\Omega \leq 0.1$ sr around the (110) axis (figure 6).

In the case of a recoil in the (100) direction, a single vacancy cannot be formed and the result strongly depends on the anion/cation mass ratio (figure 7). In KCl (mass ratio ~ 1) a transfer of the recoil momentum over great distances along (100) axis took place. As the sequence of interionic collisions changes the order of the ionic charges in the (100) atomic chain, the new configuration proved to be very instable, the ions returned to their initial positions and no metastable defect could arise (at least at the energies $E_R < 200$ eV). On the contrary, in the case of a large anion/cation mass ratio (~5.52 in NaI) cooperative collision processes took place in the (100) atomic chain accompanied by large-amplitude vibrations of light cations (figure 7).

$$\vec{r}_n = \vec{n} + \vec{u}_n / \sqrt{2}\, a_0$$

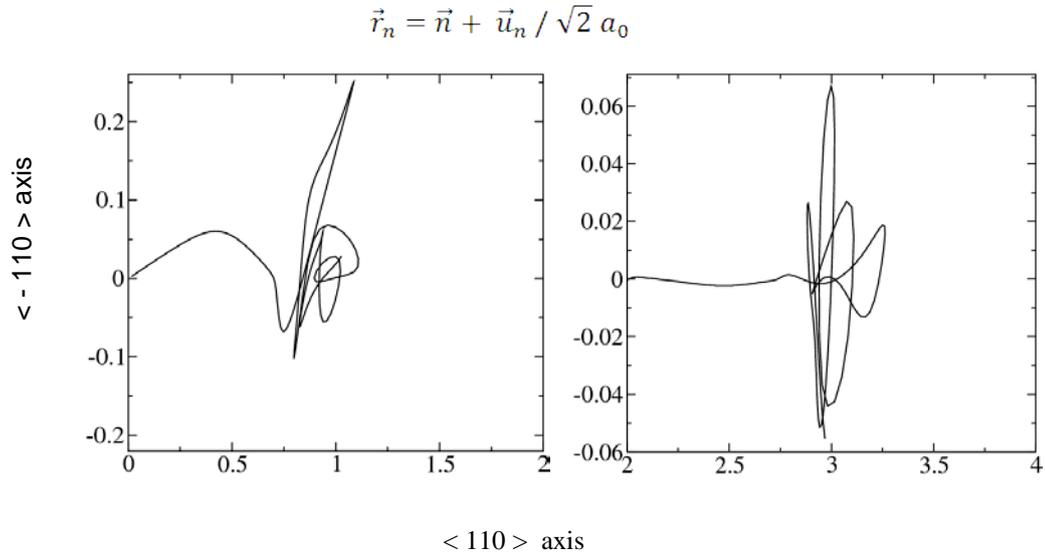

< 110 > axis

**Figure 6.** Focuson effects in NaI. The recoil momentum is directed in the (001) plane, the angle of the inclination with the (110) axis is 10°. The trajectories $\vec{r}_n$ of the primary Na ion (at left) and its second neighbour along the (110) axis (at right) are presented. Time interval is 1 ps.

As a result, a bi-vacancy near the primary ion position and an extended crowdion appeared in the chain. The crowdion was slowly drifting away from the bi-vacancy. The required minimal recoil energy exceeded almost four times the corresponding $E_R$ in the (110) direction case. According to Ref. [10] an analogous defect pair can also be created at the (111) recoil direction, the pair being more compact due to the less density of the ions in the corresponding chain.

$$r_n(t) = n + u_n(t) / a_0$$

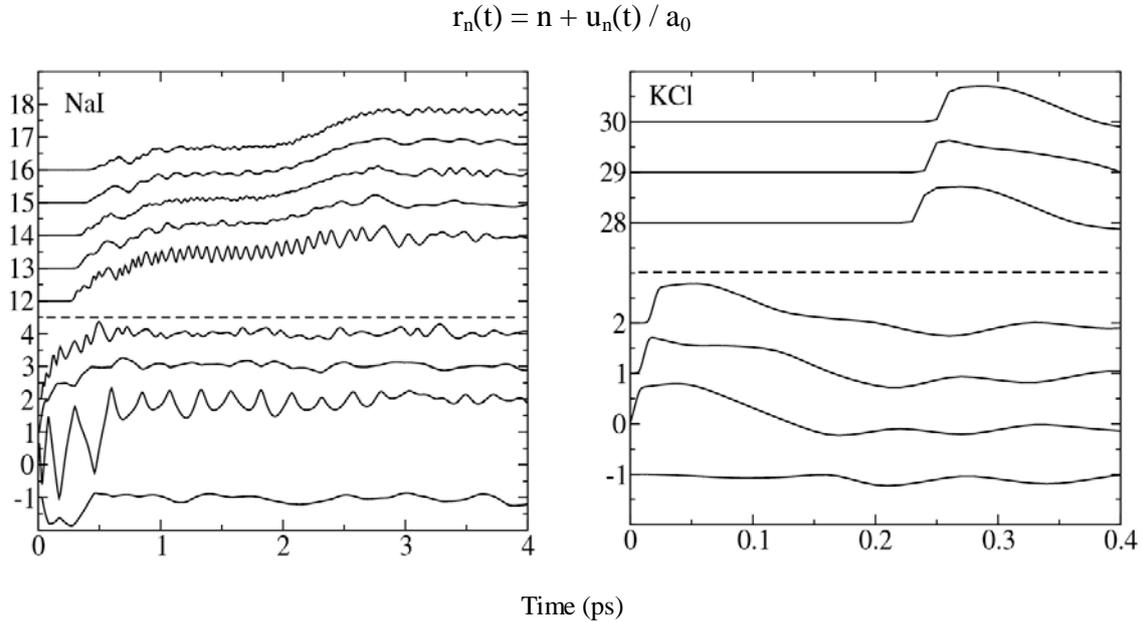

Time (ps)

**Figure 7.** Time dependence of the positions of the colliding ions in NaI and KCl crystals. Displacements are induced by a recoil momentum transferred to the primary cation ($n = 0$) in the (100) direction. Recoil energies and lattice parameters: 1) NaI: $E_R = 99$ eV, $a_0 = 3.208$ Å (at left), 2) KCl: $E_R = 190$ Å, $a_0 = 3.12$ Å (at right).

We note that the long- range forces, created by the ion displacements in the host crystal, have not been taken into account in these calculations. In our mind, their inclusion does not produce any qualitative changes, as the defect formation is mostly governed by strong short-range repulsive forces. Nevertheless, some influence on the threshold energies and on the stability of the defects is possible.

## 5. Conclusions

In this communication the results of our recent studies of the atomic motion in perfect crystals, caused by the local excitations of the crystal lattices with the energy less than 200 eV, are presented. At small energies (<10 eV) the excitation can induce intrinsic localized modes (ILMs) (see, e.g. [1-5]) and recently predicted in [6] and experimentally observed in [7] linear local modes associated with them. Usually pair potentials in dielectrics show a strong softening with the increasing of vibrational amplitudes. Therefore, in these materials, the ILMs can only split down from the optical band(s) into a spectral gap of the phonon spectrum, if such gaps exist in the crystal. However, in metals due to the screening of the ion-ion interaction by conducting electrons the ion-ion potentials may be very different as compared to the insulator cases. A checkup, provided by us, shows that in metallic Ni, Nb and Fe, due to renormalization of atomic interactions by free electrons, the ILMs with the frequencies positioned above the phonon spectrum can arise. It is shown that these ILMs are highly mobile and can efficiently transfer a concentrated vibrational energy to large distances along crystallographic directions. If the recoil energy exceeds tens of eV, the vacancies and interstitials can be formed, in a strong dependence on the direction of the recoil momentum. In the NaCl-type lattices the recoil in the (110) direction can produce a vacancy and a crowdion. However, in the case of a recoil in the (100) direction the single vacancy cannot be formed and the result strongly depends on the anion/cation mass ratio. In KCl (mass ratio ~ 1) a transfer of the recoil momentum over great distances along the (100) axis took place. As the sequence of interionic collisions changes the order of the ionic charges in the (100) atomic chain, the new configuration proved to be very instable, the ions returned to their initial positions and no metastable defect could arise (at least at the energies $E_R$ < 200 eV). On the contrary, in NaI a defect pair bi-vacancy + crowdion appears.

ACKNOWLEDGMENTS
The research was supported by Estonian research projects SF0180013s07, IUT2-27 and by the European Union through the European Regional Development Fund (project 3.2.0101.11-0029).